\documentclass[aps,prx,preprint,onecolumn,citeautoscript,superscriptaddress,nofootinbib,
eqsecnum]{revtex4-1}  
\usepackage{amsmath,amssymb,bm} 
\usepackage{graphicx}  
\usepackage{color} 
\usepackage[papersize={8.5in,11in}]{geometry}
\usepackage[colorlinks=true]{hyperref}  
\hypersetup{
    bookmarks=true,         % show bookmarks bar?
    unicode=false,          % non-Latin characters 
    pdftoolbar=true,        % show Acrobat
    pdfmenubar=true,        % show Acrobat 
    pdffitwindow=false,     % window fit to page when opened
    pdfstartview={FitH},    % fits the width of the page to the window
    pdftitle={Quantum chaos on a critical Fermi surface},    % title
    pdfauthor={Aavishkar A. Patel},     % author
    pdfsubject={},   % subject of the document
    pdfcreator={},   % creator of the document
    pdfproducer={}, % producer of the document
    pdfkeywords={} {} {}, % list of keywords
    pdfnewwindow=true,      % links in new window
    colorlinks=true,       % false: boxed links; true: colored links
    linkcolor=magenta, %red,          % color of internal links (change box color with linkbordercolor)
    citecolor=blue,        % color of links to bibliography
    filecolor=magenta,      % color of file links
    urlcolor=blue           % color of external links
} 

\usepackage{amsfonts}
\usepackage{upgreek}
\usepackage{slashed}
\usepackage{latexsym}

\usepackage{todonotes}

\newcommand{\beq}{\begin{equation}}
\newcommand{\eeq}{\end{equation}}
\def\bea{\begin{eqnarray}}
\def\eea{\end{eqnarray}}

\begin{document}

\title{Quantum chaos on a critical Fermi surface}

\author{Aavishkar A. Patel}
\affiliation{Department of Physics, Harvard University, Cambridge MA 02138, USA}

\author{Subir Sachdev}
\affiliation{Department of Physics, Harvard University, Cambridge MA 02138, USA}
\affiliation{Perimeter Institute for Theoretical Physics, Waterloo, Ontario, Canada N2L 2Y5}

\begin{abstract}
We compute parameters characterizing many-body quantum chaos for a critical Fermi surface without quasiparticle 
excitations. We examine a theory of $N$ species of fermions at non-zero density coupled to a $U(1)$ gauge field in two spatial dimensions,
and determine the Lyapunov rate and the butterfly velocity in an extended random-phase approximation. 
The thermal diffusivity is found to be universally related to these chaos parameters {\em i.e.\/} the relationship is independent of $N$, the
gauge coupling constant, the Fermi velocity, the Fermi surface curvature, and high energy details.
\end{abstract}
\maketitle

\section{Introduction}

States of quantum matter without quasiparticle excitations are expected \cite{ssbook} to have a shortest-possible local thermalization or phase coherence time
of order $\hbar /k_B T$  as $T \rightarrow 0$, where $T$ is the absolute temperature. 
Much recent attention has recently focused on the related and more precise notion of a Lyapunov time, $\tau_L$, the time to 
many-body quantum chaos \cite{Larkin1969}. By analogy with 
classical chaos, $\tau_L$ is a measure of the time
over which the wavefunction of a quantum system is scrambled by an initial perturbation. This scrambling can be measured by considering the magnitude-squared of the commutator of two observables a time $t$
apart \cite{Larkin1969,Maldacena2015}: 
the growth of the commutator with $t$ is then a measure of how the quantum state at the later
time has been perturbed since the initial observation. In chaotic systems, and with a suitable choice of observables, the growth is initially exponential, $\sim \exp (t/\tau_L)$, and this defines $\tau_L$.
With some reasonable physical assumptions on states without quasiparticles, 
it has been established that this time obeys a lower bound \cite{Maldacena2015}
\beq
\tau_L \geq \frac{\hbar}{2 \pi k_B T}; \label{mssbound}
\eeq
(henceforth, we set $k_B = \hbar =1$).
The lower bound is saturated in quantum matter states holographically dual to Einstein gravity \cite{SSDS14}, and in the SYK model
of a strange metal \cite{SY92,kitaev2015talk,Maldacena2016,Gu:2016oyy}. Relativistic theories in a vector large-$N$ limit provide a weakly-coupled realization
of states without quasiparticles, and in these cases 
$\tau_L \sim N/T$ \cite{Stanford2015,DS97,Sachdev97}, which is
larger than the bound in Eq. (\ref{mssbound}) but still of order $1/T$.
Fermi liquids have quasiparticles, and their $\tau_L \sim 1/T^2$ is parameterically
larger than Eq.~(\ref{mssbound}) as $T \rightarrow 0$ \cite{AFI16,SBEA16}.
In general we expect that $\tau_L$ is of order $1/T$ only in sytems without quasiparticle excitations.

In this paper, we turn our attention to non-Fermi liquid states of widespread interest in condensed matter physics. The canonical example we shall
examine is that of $N$ species of fermions at a non-zero density coupled to a U(1) gauge field in two spatial dimensions.
Such a theory has a Fermi surface in momentum space which survives in the presence of the gauge field\footnote{
The Fermi surface is defined by the locus of points where the inverse fermion Green's function vanishes, and is 
typically computed in the gauge $\vec{\nabla} \cdot \vec{a} = 0$: this yields the same Fermi surface
as in the closely-related problem of a Fermi surface coupled to Ising-nematic order.}, even though the fermionic quasiparticles do not.
Closely related theories apply to a wide class of problems with a critical Fermi surface, and we expect that our results can be extended to these cases too. 

It has been recognized for some time \cite{Lee2009} that the naive vector $1/N$ expansion of the critical Fermi surface problem breaks down at higher-loop orders (beyond three loops in the fermion self energy). This is in strong contrast to the behavior of relativistic theories at zero density in which the vector $1/N$ expansion is well behaved. This indicates the large $N$ theory of a critical Fermi surface is strongly-coupled. Strong-coupling effects have been examined by carefully studying higher loops, or by alternative expansion methods \cite{Metlitski2010,Mross:2010rd,Dalidovich2013,HM15}, and in the end the results are similar to those in an
random-phase approximation (RPA) 
theory \cite{Halperin1992,JP94,Furusaki1994}. So far, the main new effect discovered at strong coupling is  
a small fermion anomalous dimension, but this will not be important for our purposes here.

Here, we shall use an extended RPA theory to compute the Lyapunov time, and the associated butterfly 
velocity $v_B$ \cite{SSDS14,Roberts:2014isa,Shenker:2013yza,Blake2016,Blake:2016sud,Roberts:2016wdl,Alishahiha2014,Lucas:2016yfl}, for the critical Fermi surface in two spatial dimensions. As $T \rightarrow 0$, we find for the Lyapunov rate $\lambda_L \equiv 1/\tau_L$
\beq
\lambda_L \approx 2.48 \, T
\label{lyapunov}
\eeq
which obeys the bound $\lambda_L \leq 2 \pi T$ in Eq.~(\ref{mssbound}). Notably the value of $\lambda_L$ for the critical Fermi surface
is independent of the gauge coupling constant, $e$, and also of $N$. This supports the conclusion \cite{Lee2009} that this theory is
strongly coupled in the $N \rightarrow \infty$ limit. Our result for the butterfly velocity is more complicated; as $T \rightarrow 0$
\beq
v_B \approx 4.10 \, \frac{N T^{1/3}}{e^{4/3}}\frac{v_F^{5/3}}{\gamma^{1/3}}. \label{vbres}
\eeq
This depends upon both $N$ and $e$, and also on the Fermi velocity, $v_F$, and the Fermi surface curvature, $\gamma$.

Blake \cite{Blake2016,Blake:2016sud} has recently suggested, using holographic examples, 
that there is a universal relation between transport properties, 
as characterized by the energy and charge  diffusivities \cite{Hartnoll2015}, and 
the parameters characterizing quantum chaos $v_B$, and $\lambda_L$.
For the critical Fermi surface being studied here, momentum is conserved by the critical theory,
and so the electric conductivity is sensitive to additional perturbations which relax momentum \cite{Hartnoll:2014gba,Lucas:2016yfl}. 
However, the thermal conductivity
is well-defined and finite in the non-relativistic 
critical theory \cite{Lucas:2015pxa,Eberlein2016} even with momentum exactly conserved.
So we may define a energy diffusivity, $D^E$, which we compute building upon existing work \cite{Nave2007,Halperin1992}, and find
\beq
D^E  \approx 0.42 \frac{v_B^2}{\lambda_L}. \label{Deres}
\eeq
Notably, the factors of $e$, $N$, $v_F$ and $\gamma$ in Eq.~(\ref{vbres}) cancel precisely in the relationship Eq.~(\ref{Deres}). This supports the universality
of the relationship between thermal transport and quantum chaos in strongly-coupled states without quasiparticles.

A simple intuitive picture of this connection between chaos and transport follows from the recognizing that quantum chaos is intimately
linked to the loss of phase coherence from electron-electron interactions. As the time derivative of the local phase is determined by the local energy,
phase fluctuations and chaos are linked to interaction-induced energy fluctuations, and hence thermal transport. On the other hand, other physical ingredients
enter into the transport of other conserved charges, and so we see no reason for a universal connection between chaos and
charge transport. 

\section{Model}

\begin{figure}[h]
\begin{center}
\includegraphics[height=2.0in]{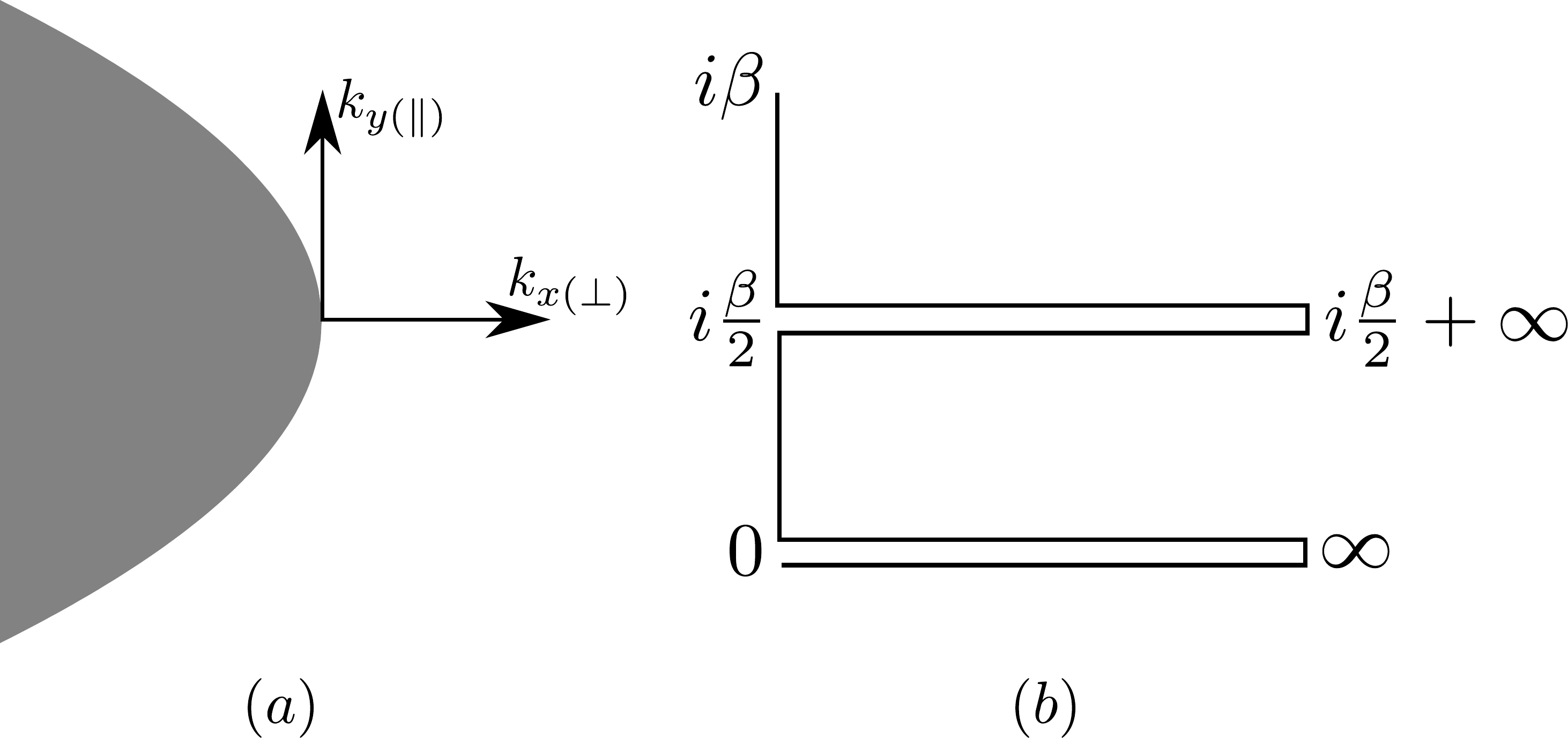}
\end{center}
\caption{(a) Fermi surface patch and coordinate system (b) The complex-time contour $C$ used for evaluating out-of-time-order correlation functions. It contains forward and backward time evolution along two real time folds separated by $i\beta/2$, and imaginary time evolution between the folds.}
\label{fsc}
\end{figure}

We consider a single patch of a Fermi surface with $N$ fermion flavors, $\psi_j$, coupled to a $U(1)$ gauge boson in two spatial dimensions: this is described by the ``chiral non-Fermi liquid" model~\cite{Sur2014} (Fig.~\ref{fsc}a). The (Euclidean) action is given by
\begin{align}
&S_e = \int \frac{d^3k}{(2\pi)^3}\left(\sum_{j=1}^N\psi_j^\dagger(k)(-i\eta k_0+\epsilon_k)\psi_j(k) + \frac{N}{2}\phi(k)(c_b|k_0|/|k_y|+k_y^2)\phi(-k)\right) \nonumber \\
&~~~~~~~~~~~~~~+ e\int \frac{d^3k}{(2\pi)^3}\sum_{j=1}^N\int \frac {d^3q}{(2\pi)^3}\phi(q)\psi_j^\dagger(k+q)\psi_j(k), \nonumber \\
&\epsilon_k = v_F k_x+ \gamma k_y^2 \quad, \quad c_b = e^2/(8\pi v_F\gamma).
\end{align}
This is derived from the action of a Fermi surface coupled to a U(1) gauge field with gauge coupling constant $e$. We only include the transverse
gauge fluctuations in the gauge $\vec{\nabla} \cdot \vec{a} = 0$, in which cause the gauge field reduces to a single boson $\phi$ representing
the component of the gauge field perpendicular to the Fermi surface. 
We have already included the one-loop boson self energy in $S_e$. Unless explicitly mentioned, we shall set the Fermi velocity $v_F$ and the Fermi surface curvature $\gamma$ to unity in the rest of this work. These factors can be restored by appropriately tracing them through the computations.  An advantage of this model is that the one-loop scaling structure of the boson and fermion Green's functions is ``exact". As there is only a single patch, the one-loop scaling structure is not destroyed by the coupling of different patches at higher loop orders~\cite{Metlitski2010}. However, this theory is still not fully controllable via the large-$N$ expansion: IR divergences in higher loop diagrams, such as the three-loop fermion self energy, enhance their coefficients by powers of $N$. Ultimately, all planar diagrams must be taken into account~\cite{Lee2009}. A version of this model that combines two antipodal patches of the Fermi surface is amenable to a more controlled $\epsilon=5/2-d$ expansion~\cite{Dalidovich2013}. However, our analysis cannot be performed easily with this dimensionally regularized construction, so we will restrict ourselves to the $d=2$ RPA theory. Despite its flaws, the RPA theory has correctly determined other physical features of this theory, such as the scaling of the optical conductivity~\cite{Furusaki1994,Eberlein2016} which agrees with 
the $\epsilon=5/2-d$ expansion \cite{Eberlein2016}.

The bare frequency dependent term in the fermion propagator is irrelevant in the scaling limit and is hence multiplied by the positive infinitesimal $\eta$. However, the presence of this term might lead to crossovers in the quantities that we compute at high temperatures. The above action is invariant under the rescaling 
\begin{align}
&k_x \rightarrow b^{-1} k_x,~~ k_y \rightarrow b^{-1/2} k_y,~~ k_0 \rightarrow b^{-3/2} k_0, \nonumber \\
&e\rightarrow e,~~\psi\rightarrow b^2\psi,~~\phi\rightarrow b^2\phi.
\end{align}
The coupling $e$ is thus dimensionless, and the dynamical critical exponent is $z=3/2$.

Since we will need to perform all computations at finite temperature, it is imperative that we understand what the finite-temperature Green's functions are. In the above patch theory, the gauge boson does not acquire a thermal mass due to gauge invariance~\cite{Hartnoll2014}. However, we will nevertheless add a very small ``mass" by hand to use as a regulator. The boson Green's function then is
\beq
D(k) = \frac{|k_y|/N}{|k_y|^3+c_b |k_0|+ m^2}.
\eeq
This boson Green's function may then be used to derive the thermally corrected fermion Green's function via the one-loop self-energy starting from free fermions~\cite{DellAnna2006} (Appendix~\ref{fse})
\beq
G(k) = \frac{1}{k_x+k_y^2-i\frac{c_f}{N}\mathrm{sgn}(k_0)T^{2/3}H_{1/3}\left(\frac{|k_0|-\pi T\mathrm{sgn}(k_0)}{2\pi T}\right)-i\mathrm{sgn}(k_0)\frac{\mu(T)}{N}},~~\mu(T)=\frac{e^2T}{3\sqrt{3}m^{2/3}},
\eeq
($c_f=2^{5/3}e^{4/3}/(3\sqrt{3})$)
where $\mu(T)$ is generated by $m^2$ cutting off an IR divergence coming from the zeroth boson Matsubara frequency, and $H_{1/3}(x)$ is the analytically continued harmonic number of order $1/3$, with
\beq
H_{r}(n\in\mathbb Z^+) \equiv \sum_{j=1}^{n}\frac{1}{j^{r}},~~H_{r}(z) = \zeta(r) - \zeta(r, z+1).
\eeq 
This thermally corrected Green's function is not exact owing to the uncontrolled nature of the large-$N$ expansion. Higher (three and beyond) loop corrections to the fermion self energy also contain terms that are ultimately of order $1/N$, which will modify the self energy but should leave the relative scalings of frequency, momentum and temperature unchanged~\cite{Lee2009}. The same is also true for various other diagrams. As such, the numerical prefactors in the Lyapunov exponent and butterfly velocity that we determine may not be exact, but we should be able to correctly deduce their scaling properties.

\section{Scrambling and the Lyapunov exponent}

To study out-of-time-order correlation functions, we define the path integral on a contour $C$ which runs along both the real and imaginary time directions, with two real-time folds separated by $i\beta/2$~\cite{Stanford2015, Maldacena2016} (Fig.~\ref{fsc}b). The generating functional is given by
\beq
\mathcal{Z} = \int_C \mathcal{D}\bar{\psi}\mathcal{D}\psi\mathcal{D}\phi e^{i S[\bar{\psi},\psi,\phi]}.
\eeq

To measure scrambling, we will use fermionic operators, and hence we replace the commutators \cite{Larkin1969} by anti-commutators.
We will evaluate the index-averaged squared anticommutator~\cite{Stanford2015, Maldacena2016}
\beq
f(t) = \frac{1}{N^2}\theta(t)\sum_{i,j=1}^{N}\int d^2x~\mathrm{Tr}\left[e^{-\beta H/2}\{\psi_i(x,t),\psi_j^\dagger(0)\}e^{-\beta H/2}\{\psi_i(x,t),\psi_j^\dagger(0)\}^\dagger\right] =\int d^2x~f(t,x). 
\label{ft}
\eeq
This function is real and invariant under local $U(1)$ gauge transformations of the $\psi$s. The staggered factors of $e^{-\beta H/2}$ place two of the field operators on each of the real time folds. $f(t)$ contains the out-of-time-ordered correlation function $\langle\psi(x,t)\psi^\dagger(0)\psi^\dagger(x,t)\psi(0)\rangle$ that describes scrambling. The anticommutators simplify the evaluation in comparison to the correlation function of just the four fermionic operators. $f(t)$ then measures how the operators ``spread" as a function of time. At $t = 0$, the anticommutators vanish for $x \neq 0$. At later times, the operators become non-local under the time 
evolution, leading to a growth of the function. It is conjectured~\cite{Maldacena2015} that at short times
\beq
f(t) \sim e^{\lambda_L t}+~...~, 
\label{ftexp}
\eeq
where $0\le\lambda_L\le 2\pi T$ is the Lyapunov exponent. Our goal is to compute $\lambda_L$.  At long times, which we are not interested in, $f(t)$ saturates to some finite asymptotic value. Formally, to precisely define $\lambda_L$, we need the growing exponential
in (\ref{ftexp}) to have a small prefactor. This can be provided here by examining spatially separated correlators (which we
shall do in Section~\ref{sec:butterfly}), although not by the large $N$ limit. Operationally, for now, we will compute $f(t)$ by
using diagrams similar to those employed in relativistic theories \cite{Stanford2015}.

\begin{figure}
\begin{center}
\includegraphics[width=\textwidth]{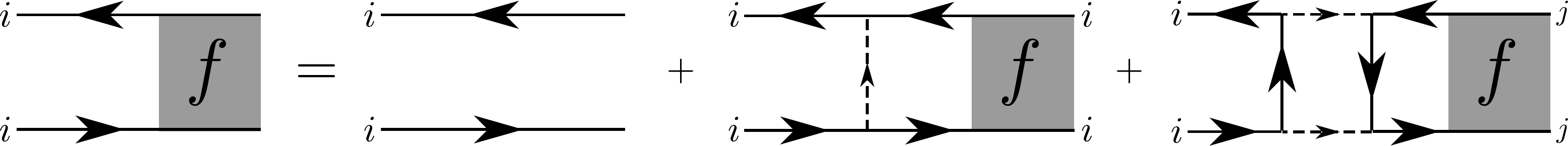}
\end{center}
\caption{The Bethe-Salpeter equation for $f(\omega)$ at leading naive order in $1/N$. Solid lines are fermion propagators and dashed lines are boson propagators. The arrows indicate the directions of momentum flow used in the equations in the text. For the fermion lines, advanced Green's functions are used for the upper rails and retarded ones for the lower rails, as can be seen from Eq.~(\ref{ft}). The third diagram on the right hand side is the same order in $1/N$ as the second despite having two boson propagators, because it involves summing over the flavors $j$.}
\label{bs}
\end{figure}

The approach described in Ref.~\cite{Stanford2015} involves summing a series of diagrams to obtain $f(t)$. The simplest subset of these is a ladder series (Fig.~\ref{bs}), with the ``rails" of the ladder defined on the real-time folds, and the ``rungs" connecting times separated by $i\beta/2$. The interaction vertices are integrated only over the real-time folds as an approximation to minimize technical complexity; more general placements are expected to make corrections to the thermal state that should not affect $\lambda_L$. The end result is that one uses retarded Green's functions for the rails (since the real time folds involve both forward and backward evolution) and Wightman functions for the rungs~\cite{Stanford2015}:
\begin{align}
&G^R(x,t) = -i\theta(t)\mathrm{Tr}\left[e^{-\beta H}\{\psi(x,t),\psi^\dagger(0)\}\right],  \nonumber \\
&G^R(k) = \frac{1}{k_x+k_y^2-i\frac{c_f}{N}T^{2/3}H_{1/3}\left(-\frac{i k_0+\pi T}{2\pi T}\right)-i\frac{\mu(T)}{N}}, \nonumber \\
&G^W(x,t) = \mathrm{Tr}\left[e^{-\beta H}\psi(x,t)\psi^\dagger(0,i\beta/2)\right] = \mathrm{Tr}\left[e^{-\beta H/2}\psi(x,t)e^{-\beta H/2}\psi^\dagger(0)\right], \nonumber \\
&G^W(k) = \frac{A(k)}{2\cosh{\frac{\beta k_0}{2}}}, \nonumber \\
&D^R(x,t) = -i\theta(t)\mathrm{Tr}\left(e^{-\beta H}[\phi(x,t),\phi(0)]\right), \nonumber \\
&D^R(k) = \frac{|k_y|/N}{|k_y|^3-ic_bk_0+m^2}, \nonumber \\
&D^W(x,t) = \mathrm{Tr}\left[e^{-\beta H}\phi(x,t)\phi(0,i\beta/2)\right] = \mathrm{Tr}\left[e^{-\beta H/2}\phi(x,t)e^{-\beta H/2}\phi(0)\right], \nonumber \\
&D^W(k) = \frac{B(k)}{2\sinh{\frac{\beta k_0}{2}}} = \frac{1}{N}\frac{c_b k_0}{\sinh{\frac{\beta k_0}{2}}} \frac{|k_y|}{(|k_y|^3+m^2)^2+c_b^2k_0^2}.
\label{green}
\end{align}
($A$ is the fermion spectral function and $B$ is the boson spectral function).
For an explicit derivation of the Wightman functions see Appendix~\ref{wight}. There are two types of rungs at leading order in $1/N$: one is simply the boson Wightman function. The other is  a ``box" that contains fermion Wightman functions and retarded boson functions. 

The first diagram in the ladder series which has no rungs is given by
\begin{align}
&f_0(t) = \frac{1}{N}\int d^2x~|G^R(x,t)|^2, \nonumber \\
&f_0(\omega) = \frac{1}{N}\int\frac{d^3k}{(2\pi)^3}G^R(k)G^{R\ast}(k-\omega) \nonumber \\
&= \frac{i}{N} \int\frac{dk_ydk_0}{(2\pi)^2}\frac{1}{i\frac{c_f}{N}T^{2/3}\left[H_{1/3}\left(-\frac{ik_0+\pi T}{2\pi T}\right)+H_{1/3}\left(-\frac{i(\omega-k_0)+\pi T}{2\pi T}\right)\right]+2i\frac{\mu(T)}{N}}.
\label{f0}
\end{align}
This bare term remarkably ends up being $\mathcal{O}(1)$. Since $m$ is tiny, $\mu(T)\rightarrow+\infty$. In the time domain, this thus describes a function that decays exponentially very quickly.

We have the Bethe-Salpeter equation of the ladder series
\begin{align}
&f(\omega) = \frac{1}{N} \int \frac{d^3k}{(2\pi)^3} f(\omega,k) \nonumber \\
&= \frac{1}{N} \int \frac{d^3k}{(2\pi)^3} G^R(k)G^{R\ast}(k-\omega)\left[1+ \int \frac{d^3k^\prime}{(2\pi)^3}\left(e^2 D^W(k-k^\prime) + K_2(k,k^\prime,\omega)\right)f(\omega,k^\prime)\right], \nonumber \\
&f(\omega,k) = G^R(k)G^{R\ast}(k-\omega)\left[1+ \int \frac{d^3k^\prime}{(2\pi)^3}\left(e^2 D^W(k-k^\prime) + K_2(k,k^\prime,\omega)\right)f(\omega,k^\prime)\right],
\label{BS}
\end{align}
The sign of the $e^2D^W$ term is $+1$, coming from a factor of $-i^2$ arising from the $i$'s in in the interaction vertex of $S$. We need to solve this integral equation to determine the behavior of $f(t)$. We note that as in Ref.~\cite{Stanford2015}, the condition for $f(t)$ to grow exponentially is that the ladder sum be invariant under the addition of an extra unit to the ladder, i.e. 
\beq
f(\omega,k) = G^R(k)G^{R\ast}(k-\omega)\int \frac{d^3k^\prime}{(2\pi)^3}\left(e^2 D^W(k-k^\prime) + K_2(k,k^\prime,\omega)\right)f(\omega,k^\prime),
\eeq

We have
\beq
K_2(k,k^\prime,\omega) = N e^4 \int \frac{d^3k_1}{(2\pi)^3}D^R(k_1)D^{R\ast}(k_1-\omega)G^W_0(k-k_1)G^W_0(k^\prime-k_1).
\eeq
The overall sign of this contribution is also $i^2(-i)^2=1$, where the factors of $i$ again come from the four interaction vertices. Here we use the bare fermion Wightman functions, as the self energy corrections will come in at higher orders in $1/N$. As the integral is free of IR divergences, the overall power of $1/N$ in this contribution is not enhanced and this simplification should be safe. In the bare fermion Wightman functions, we drop the frequency dependent term that is irrelevant at low frequencies by sending $\eta\rightarrow0$, to preserve the quantum critical scaling
\beq
G_0^W(k) = \frac{A(k)}{2\cosh\frac{\beta k_0}{2}} \rightarrow \frac{\pi\delta(k_x+k_y^2)}{\cosh\frac{\beta k_0}{2}}.
\eeq
(A is the fermion spectral function). There is a $\cosh$ instead of a $\sinh$ in the fermion Wightman function (Appendix~\ref{wight}). We then have
\begin{align}
&K_2(k,k^\prime,\omega) = \frac{e^4}{N} \int \frac{d^3k_1}{(2\pi)^3}\frac{k_{1y}^2}{(|k_{1y}|^3-ic_bk_{10}+m^2)(|k_{1y}|^3+ic_b(k_{10}-\omega)+m^2)}\nonumber \\ 
&\times\frac{\pi^2\delta(k_x-k_{1x}+(k_y-k_{1y})^2)\delta(k^\prime_x-k_{1x}+(k^\prime_y-k_{1y})^2)}{\cosh\frac{k_0-k_{10}}{2T}\cosh\frac{k^\prime_0-k_{10}}{2T}}.
\end{align}

Since there are no IR divergences, we drop the $m^2s$. Doing the $k_{1x}$ integral followed by the $k_{1y}$ one, this simplifies to
\begin{align}
&K_2(k,k^\prime,\omega) = \frac{e^4}{\pi N} \int dk_{10} \frac{(\epsilon_k-\epsilon_{k^\prime})^2 |k_y-k^\prime_y|^3}{(|\epsilon_k-\epsilon_{k^\prime}|^3-8ic_bk_{10}|k_y-k^\prime_y|^3)(|\epsilon_k-\epsilon_{k^\prime}|^3+8ic_b(k_{10}-\omega)|k_y-k^\prime_y|^3)}\nonumber \\ 
&\times\frac{1}{\cosh\frac{k_0-k_{10}}{2T}\cosh\frac{k^\prime_0-k_{10}}{2T}}.
\end{align}
Due to the sliding symmetry along the Fermi surface~\cite{Metlitski2010}, we expect the eigenfunction that we are interested in to obey $f(\omega,k)=f(\omega,k_0,\epsilon_k)$. This can be proven by induction considering the series of diagrams that we sum. We can then shift $k^\prime_x\rightarrow k^\prime_x-k_y^{\prime2}$ followed by $k^\prime_y\rightarrow k^\prime_y+k_y$ and integrate over $k^\prime_y$
\begin{align}
&\int \frac{d^3k^\prime}{(2\pi)^3}K_2(k,k^\prime,\omega)f(\omega,k^\prime)= \frac{e^4}{24\pi\sqrt{3}c_b^{4/3}N}\int \frac{dk^\prime_0dk^\prime_x}{(2\pi)^2} \int dk_{10} \frac{k_{10} \left((-i k_{10})^{1/3}-(i (k_{10}-\omega ))^{1/3}\right)}{(-i k_{10})^{4/3} (i (k_{10}-\omega ))^{1/3} (2 k_{10}-\omega )} \nonumber \\
&\times\frac{f(\omega,k^\prime_0,k^\prime_x)}{\cosh\frac{k_0-k_{10}}{2T}\cosh\frac{k^\prime_0-k_{10}}{2T}},
\end{align}
and
\begin{align}
&e^2 \int \frac{d^3k^\prime}{(2\pi)^3}D^W(k-k^\prime)f(\omega,k^\prime) \nonumber \\
&= \frac{e^2}{N} \left[\lim_{m\rightarrow0} \int \frac{d^3k^\prime}{(2\pi)^3}\left(|k_y^\prime-k_y|\frac{\tilde{f}(\omega,k_0^\prime,k^\prime_x)\frac{c_b(k_0^\prime-k_0)}{\sinh\frac{\beta(k_0^\prime-k_0)}{2}}-2c_bT\tilde{f}(\omega,k_0,k^\prime_x)}{(|k_y^\prime-k_y|^3+m^2)^2+c_b^2(k_0^\prime-k_0)^2}\right)+\frac{2\mu(T)}{e^2}\right],
\end{align}
where we added and subtracted terms to make the IR divergences explicit. If we expand the numerator of the integrand in the above for $k_0^\prime\rightarrow k_0$, we see that the integral is finite and free of IR divergences.

Interestingly, both pieces of the kernel no longer depend on $k_x$ and $k^\prime_x$. Thus we can integrate both sides of the equation over $k_x$ and $k^\prime_x$ to get an equation for $\tilde{f}(\omega,k_0) \equiv \int \frac{dk_x}{2\pi} f(\omega,k_0,k_x)$. From Eq.~(\ref{f0}), we can see that the IR divergent piece $\propto\mu(T)$ cancels out. The dependence on $N$ also cancels out. We finally get 
\begin{align}
&e^2\lim_{m\rightarrow0}\int \frac{dk_0^\prime dk_y^\prime}{(2\pi)^2}|k_y^\prime|\frac{\tilde{f}(\omega,k_0^\prime)\frac{c_b(k_0^\prime-k_0)}{\sinh\frac{\beta(k_0^\prime-k_0)}{2}}-2c_bT\tilde{f}(\omega,k_0)}{(|k_y^\prime|^3+m^2)^2+c_b^2(k_0^\prime-k_0)^2} \nonumber \\
&+ \frac{e^4}{24\pi\sqrt{3}c_b^{4/3}}\int \frac{dk^\prime_0dk_{10}}{2\pi}\frac{k_{10} \left((-i k_{10})^{1/3}-(i (k_{10}-\omega ))^{1/3}\right)}{(-i k_{10})^{4/3} (i (k_{10}-\omega ))^{1/3} (2 k_{10}-\omega)}\frac{\tilde{f}(\omega,k^\prime_0)}{\cosh\frac{k_0-k_{10}}{2T}\cosh\frac{k^\prime_0-k_{10}}{2T}} \nonumber \\
&=c_fT^{2/3}\left(H_{1/3}\left(-\frac{i k_0 +\pi T}{2\pi T}\right)+H_{1/3}\left(-\frac{i (\omega-k_0) +\pi T}{2\pi T}\right)\right)\tilde{f}(\omega,k_0).
\label{iefinal}
\end{align}
As a matrix equation, this is of the form $M(\omega)\tilde{f}(\omega)=0$. Since we are looking for a positive growth exponent, we need to numerically find solutions of this equation on the positive imaginary $\omega$ axis. The analytic continuations of the self-energies that we made are still valid as long as $\mathrm{Im}[\omega]>0$. The largest solution will provide the growth exponent $\lambda_L$. We can see from the above equation and from the quantum critical scaling $k_0,k_0^\prime \sim T$,~~$k_y,k_y^\prime \sim e^{2/3}T^{1/3}$ that $\lambda_L\propto T$ and is independent of $e$. The numerical solution to this equation is detailed in Appendix~\ref{nume}. We find that $\lambda_L \approx 2.48~T$, which is well within the bound of Ref.~\cite{Maldacena2015}. We further see that $\lambda_L$ is not suppressed by powers of $1/N$, unlike other vector models in the large-$N$ limit. This indicates that this theory is strongly coupled at the lowest energy scales, even for large values of $N$. 

The cancellation of the IR divergent piece between the self-energy and ladder diagrams has an important physical meaning. Besides being required by gauge invariance (as $f(t,x)$ is gauge invariant), it indicates that ``classical" processes do not contribute to many-body quantum chaos: The IR divergent terms arise from classical collisions in which the frequency of the boson is zero. In this limit, the boson behaves like a thermally (but not quantum-mechanically) fluctuating random potential for the fermions, each instance of which can be described by an integrable quadratic Hamiltonian, and is hence unable to induce chaos. 

At high temperatures, when $NT^{1/3}/e^{4/3}\gg 1$, we may no longer be able to neglect the bare frequency dependent term in the fermion propagators. This would essentially amount to adding a term $\sim N\omega\tilde{f}(\omega,k_0)$ to the right hand side of Eq.~(\ref{iefinal}). Counting powers, we then might expect $\lambda_L\sim e^{4/3}T^{2/3}/N$. In Appendix~\ref{hoc} we consider a few higher order (in $1/N$) corrections to the ladder series and show that some of them are insignificant.

\section{The butterfly effect and energy diffusion}
\label{sec:butterfly}
\subsection{Butterfly velocity}
The out-of-time-order correlation function evaluated at spatially separated points characterizes the divergence of phase space trajectories in both space and time. This process is described by the function $f(t,x)$ defined in Eq.~(\ref{ft}), which is the same as the function $f(t)$ we used to determine $\lambda_L$ except for the integration over spatial coordinates. This function should contain a traveling wave term that propagates with a speed known as the ``butterfly velocity"~\cite{Roberts:2016wdl}.  In order to compute this function we will need to evaluate the ladder diagrams at a finite external momentum $p$. For simplicity, and since the component of the Fermi velocity perpendicular to the Fermi surface dominates the one parallel to the Fermi surface, we will take the external momentum to also be perpendicular to the Fermi surface. This will allow us to determine the component of the butterfly velocity perpendicular to the Fermi surface ($v_{B\perp}$). 

Repeating the same steps that led to the derivation of Eq.~(\ref{iefinal}), we simply obtain
\begin{align}
&e^2\lim_{m\rightarrow0}\int \frac{dk_0^\prime dk_y^\prime}{(2\pi)^2}|k_y^\prime|\frac{\tilde{f}(p_x,\omega,k_0^\prime)\frac{c_b(k_0^\prime-k_0)}{\sinh\frac{\beta(k_0^\prime-k_0)}{2}}-2c_bT\tilde{f}(p_x,\omega,k_0)}{(|k_y^\prime|^3+m^2)^2+c_b^2(k_0^\prime-k_0)^2} \nonumber \\
&+ \frac{e^4}{24\pi\sqrt{3}c_b^{4/3}}\int \frac{dk^\prime_0dk_{10}}{2\pi}\frac{k_{10} \left((-i k_{10})^{1/3}-(i (k_{10}-\omega ))^{1/3}\right)}{(-i k_{10})^{4/3} (i (k_{10}-\omega ))^{1/3} (2 k_{10}-\omega)}\frac{\tilde{f}(p_x,\omega,k^\prime_0)}{\cosh\frac{k_0-k_{10}}{2T}\cosh\frac{k^\prime_0-k_{10}}{2T}} \nonumber \\
&=c_fT^{2/3}\left(H_{1/3}\left(-\frac{i k_0 +\pi T}{2\pi T}\right)+H_{1/3}\left(-\frac{i (\omega-k_0) +\pi T}{2\pi T}\right)\right)\tilde{f}(p_x,\omega,k_0)+iNp_x\tilde{f}(p_x,\omega,k_0).
\label{iefinalb}
\end{align}
For small $p_x$, we expect the change in exponent ${\delta\lambda_L}/{T}\sim -{iNp_x}/({e^{4/3}T^{2/3}})$. This implies that
\beq
\int dy~f(t,x) \sim e^{\lambda_L t} \int \frac{dp_x}{2\pi} g(t,Np_x) e^{ip_x(x-v_{B\perp}t)},~~v_{B\perp}\sim \frac{NT^{1/3}}{e^{4/3}}.
\label{wave}
\eeq 
The structure of the above equation indicates that chaos propagates as a wave pulse that travels at the butterfly velocity. The wave pulse is not a soliton and broadens as it moves \cite{Roberts:2016wdl}; this is encoded in the function $g(t,Np_x)$ and further details are provided in Appendix~\ref{nume}. Note that this shows $v_{B\perp}\sim T^{1-1/z}$, which can also be straightforwardly derived by using the appropriate scalings of space and time, $[x] = -1$ and $[t] = -z$, and is also seen in holographic models~\cite{Blake2016}. Numerically we find that ${\delta\lambda_L}/{T}\approx -4.10({iNp_x}/({e^{4/3}T^{2/3}}))$, giving the result of Eq.~(\ref{vbres}) once the factors of Fermi velocity $v_F$ and Fermi surface curvature $\gamma$ are restored. (Appendix~\ref{nume}). 

This is again strictly valid only at the lowest temperatures, where $NT^{1/3}/e^{4/3} \ll 1$. Thus the butterfly velocity cannot be arbitrarily large in the large-$N$ limit. When $NT^{1/3}/e^{4/3}\sim 1$, the structure of the fermion propagator indicates that there may be a crossover to a $z=1$ regime, in which $v_{B\perp}$ will become a constant independent of $N$ and $T$. 

With the scalings $[y]=-1/2$ and $[t]=-z$, we see that the component parallel to the Fermi surface, $v_{B\parallel}\sim T^{2/3}$, which is smaller than $v_{B\perp}$ at low temperatures. Then the butterfly effect will be dominated by propagation perpendicular to the Fermi surface in the scaling limit.

\subsection{Energy diffusion}
It has been conjectured, and shown in holographic models~\cite{Hartnoll2015,Blake2016} that the butterfly effect controls diffusive transport. The thermal diffusivity 
\beq
D^E = \frac{\kappa}{C_V} \sim \frac{v_B^2}{2\pi T},
\label{dcon}
\eeq
where $\kappa$ is the thermal conductivity and $C_V$ is the specific heat at fixed density. In holographic theories $\lambda_L=2\pi T$, so a more appropriate phrasing of the above equation is $D^E\sim v_B^2/\lambda_L$~\cite{Swingle2016}. We can compute $C_V$ using the free energy of the fermions (the contribution of the boson is expected to be subleading at low temperatures~\cite{Eberlein2016}) 
\beq
F = - NT\sum_{k_0}\int\frac{d^2k}{(2\pi)^2}\ln \tilde{G}^{-1}(k),~~C_V = -T\frac{\partial^2 F}{\partial T^2},
\eeq 
where we use the one-loop dressed fermion propagator at zero temperature~\cite{Metlitski2010},
\beq
\tilde{G}^{-1}(k) = k_x + k_y^2 - i\frac{\tilde{c}_f}{N}\mathrm{sgn}(k_0)|k_0|^{2/3},~~\tilde{c}_f=\frac{3c_f}{2(2\pi)^{2/3}}.
\eeq
This computation is carried out in Appendix~\ref{shtc}. We obtain
\beq
C_V = \frac{10(2^{2/3}-1)}{9(2\pi)^{2/3}}\Gamma(5/3)\zeta(5/3)T^{2/3}e^{4/3}\frac{\gamma^{1/3}}{v_F^{2/3}}\int\frac{dk_y}{2\pi},
\label{spht}
\eeq
where we have again restored $v_F$ and $\gamma$.

Since the theory of a single Fermi surface patch is chiral, currents are non-zero even in equilibrium. We must thus define conductivities with respect to the additional change in these currents when electric fields and temperature gradients are applied. The thermal conductivity $\kappa$ is finite in the DC limit as it is defined under conditions where no additional electrical current flows. This can be achieved by simultaneously applying an electric field and a temperature gradient such that there is only an additional energy current but no additional electrical current. We have
\beq
\kappa = \bar{\kappa} - \frac{\alpha^2T}{\sigma},
\eeq
where $\alpha,\bar{\kappa}$ are the thermoelectric conductivities and $\sigma$ is the electrical conductivity. The infinities in the DC limit cancel between $\bar{\kappa}$ and the other term, yielding a finite $\kappa$~\cite{Eberlein2016}. $\bar{\kappa}$ may be obtained from the Kubo formula~\cite{Moreno1996}
\beq
\bar{\kappa}_{\perp} = -\beta\mathrm{Re}\left[\lim_{\omega\rightarrow0}\frac{\partial}{\partial\omega}i\langle J^E_\perp J^E_\perp \rangle(iq_0\rightarrow\omega+i0^+)\right],
\eeq
with the energy current
\begin{align}
&J^E_\perp(iq_0) = -i\int\frac{d^3k}{(2\pi)^3}\left(k_0+\frac{q_0}{2}\right)\frac{\partial \epsilon_k}{\partial k_x}\psi^\dagger(k+q_0)\psi(k) \nonumber \\
&= -i\int\frac{d^3k}{(2\pi)^3}\left(k_0+\frac{q_0}{2}\right)\psi^\dagger(k+q_0)\psi(k).
\end{align}
We compute the conductivities using the one-loop dressed fermion propagators in Appendix~\ref{shtc} (The boson again does not contribute directly due to the absence of an $x$-dependent term in its dispersion). The simplest vertex correction vanishes due to the structure of the fermion dispersions and other corrections are in general suppressed by powers of $N$. In this approximation $\bar{\kappa}_\perp$ is finite and $\alpha_{\perp}\propto\langle J^E_\perp J_\perp\rangle$ (where $J_\perp$ is the charge current) vanishes, so $\kappa_\perp = \bar{\kappa}_\perp$. Note that, in reality, $\bar{\kappa},\alpha,\sigma$ would all be infinite and their combination into $\kappa$ would be finite, but the final finite value of $\kappa$ should be qualitatively similar to the value obtained from our approximation. We obtain (restoring $v_F$ and $\gamma$)
\beq
\kappa_{\perp} \approx 0.28\frac{N^2T^{1/3}}{\tilde{c}_f}\frac{v_F^{8/3}}{\gamma^{1/3}}\int\frac{dk_y}{2\pi}.
\label{thc}
\eeq
Using Eqs.~(\ref{lyapunov}),~(\ref{vbres}),~(\ref{dcon}),~(\ref{spht}) we then see that
\beq
D^E_{\perp} \approx  2.83 \frac{N^2}{e^{8/3}T^{1/3}}\frac{v_F^{10/3}}{\gamma^{2/3}} \approx 0.42 \frac{v_{B\perp}^2}{\lambda_L}.
\eeq
The factors of powers of $T,N$ and $e$ match exactly on both sides of the equation and the constant of proportionality between $D^E_{\perp}$ and $v_{B\perp}^2/\lambda_L$ is an $\mathcal{O}(1)$ number. This strongly indicates that the butterfly effect is responsible for diffusive energy transport in this theory. The DC electrical conductivity is however infinite due to translational invariance, and hence, unfortunately, such a statement cannot be made for charge transport in this model. Note that the hyperscaling violating factor $\int\frac{dk_y}{2\pi}$~\cite{Eberlein2016} cancels between $\kappa_\perp$ and $C_V$. However, if we consider $\kappa_\parallel$, this does not happen due to the additional $k_y$ dependence in $J^E_\parallel$. Thus $D^E_{\parallel}$ will not be given by $v_{B\parallel}^2/\lambda_L$.

\section{Discussion}
We have computed the Lyapunov exponent $\lambda_L$ and butterfly velocity $v_B$ for a single patch of a Fermi surface with $N$ fermion flavors coupled to a $U(1)$ gauge field. At the lowest energy scales, this theory is strongly coupled regardless of the value of $N$, and we hence find that $\lambda_L$ is independent of $N$ to leading order in $1/N$. The proposed universal bound of $\lambda_L\le2\pi T$ is also obeyed. While the $1/N$ expansion is not fully controllable, it has nevertheless been capable of correctly determining many physical features of this theory in the past. We find that the butterfly velocity is dominated by propagation perpendicular to the Fermi surface, and that $v_{B\perp}\sim NT^{1/3}$. Most interestingly, we find that the butterfly effect controls diffusive transport in this model, with the thermal diffusivity $D^E_{\perp}\propto v_{B\perp}^2/\lambda_L$. Our results are valid at the lowest energy scales, at which the quantum critical scaling holds. At high temperatures, we might expect $\lambda_L$ to cross over to a slower $T^{2/3}/N$ scaling, and that $v_{B\perp}$ simply becomes a constant independent of $N$ and $T$. While technically much more complex to obtain, it would be interesting to compare the results derived from a more controlled calculation, such as the $\epsilon=5/2-d$ expansion for the two-patch version of the problem, with our results. Finally, we note recent experimental measurements of thermal diffusivity in the cuprates \cite{Aharon2017} which find a strong coupling to phonons. It would be of interest to extend the chaos theories to include the electron-phonon coupling.

\section*{Acknowledgements}
We thank M. Blake, D. Chowdhury, R. Davison,  A. Eberlein, D. Stanford and B. Swingle for valuable discussions. This research was supported by the NSF under Grant DMR-1360789 and the MURI grant W911NF-14-1-0003 from ARO. Research at Perimeter Institute is supported by the Government of Canada through Industry Canada and by the Province of Ontario through the Ministry of Research and Innovation. SS also acknowledges support from Cenovus Energy at Perimeter Institute.

\appendix

\section{Self energies}
\label{fse}

\begin{figure}[h]
\begin{center}
\includegraphics[height=1.0in]{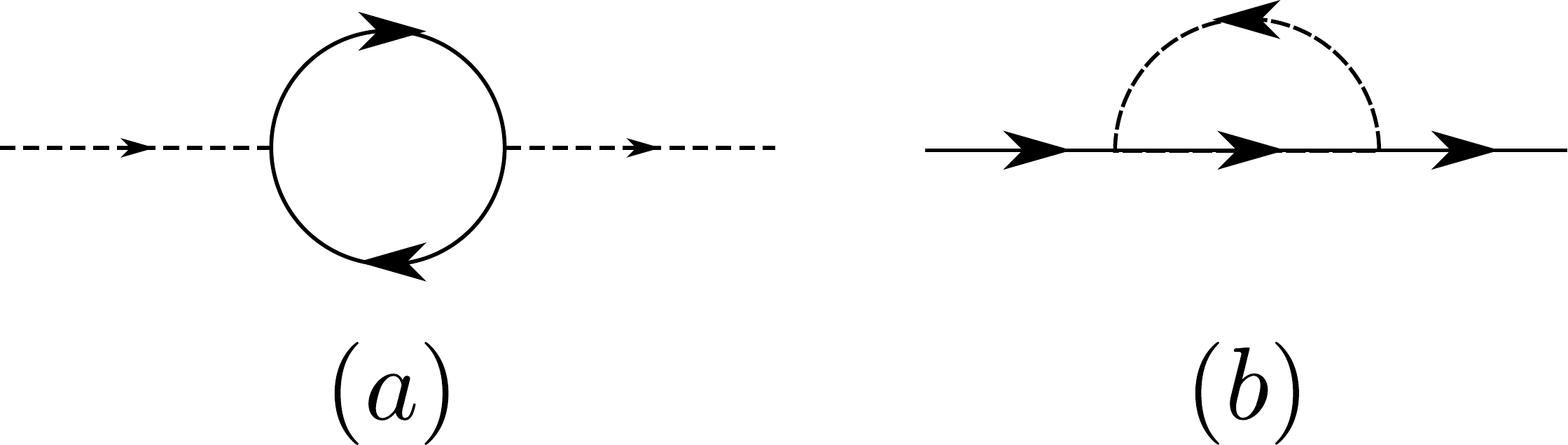}
\end{center}
\caption{(a) The one-loop boson and (b) fermion self energies. These graphs are evaluated at a finite temperature. The dashed lines are bare boson propagators and solid lines are bare fermion propagators. The arrows indicate the directions of momentum flow used in the equations in the text.}
\label{ses}
\end{figure}

The one-loop self energy graphs are shown in Fig.~\ref{ses}. The derivation of the one-loop boson self energy is standard~\cite{Metlitski2010}
\begin{align}
&\Pi(k) = -Ne^2T\sum_{q_0}\int\frac{d^2q}{(2\pi)^2} \frac{1}{q_x+q_y^2-iq_0}\frac{1}{(k_x+q_x)+(k_y+q_y)^2-i(q_0+k_0)} \nonumber \\
&=-Ne^2\int\frac{d^2q}{(2\pi)^2}\frac{n_f(q_x+q_y^2)-n_f(k_x+q_x+(k_y+q_y)^2)}{q_y^2-(k_y+q_y)^2+ik_0-k_x} \nonumber \\
&=\frac{Ne^2}{2|k_y|}\int\frac{dq_y}{(2\pi)^2}\frac{q_y^2}{q_y^2+k_0^2} =-\frac{Ne^2|k_0|}{8\pi|k_y|}+\Pi_\infty
\end{align}
The formally infinite piece $\Pi_\infty$ is tuned away by the mass renormalization at the critical point, giving the expression for the boson propagator in the main text.

The one loop fermion self energy is given by
\begin{align}
&\Sigma(k) = \frac{e^2}{N}T\sum_{q_0}\int\frac{d^2q}{(2\pi)^2}\frac{|q_y|}{|q_y|^3+c_b|q_0|+m^2}\frac{1}{k_x+q_x+(k_y+q_y)^2-i(k_0+q_0)} \nonumber \\
&=\frac{ie^2}{2N}T\sum_{q_0}\int\frac{dq_y}{2\pi}\frac{|q_y|}{|q_y|^3+c_b|q_0|+m^2}\mathrm{sgn}(k_0+q_0) \nonumber \\
&=\frac{ie^2}{3\sqrt{3}c_b^{1/3}N}T\sum_{q_0\neq0}\frac{\mathrm{sgn}(k_0+q_0)}{|q_0|^{1/3}} + i\mathrm{sgn}(k_0)\frac{e^2T}{3\sqrt{3}m^{2/3}N} \nonumber \\
&=\frac{2ie^2\mathrm{sgn}(k_0)}{3\sqrt{3}c_b^{1/3}(2\pi)^{1/3}N}T^{2/3}\sum_{n_q=1}^{|n_k|}\frac{1}{n_q^{1/3}}+i\mathrm{sgn}(k_0)\frac{\mu(T)}{N}~~~(k_0 = 2\pi T(n_k+1/2),~q_0 = 2\pi Tn_q) \nonumber \\
&=i\frac{c_f\mathrm{sgn}(k_0)}{N}T^{2/3}H_{1/3}(|n_k|)+i\mathrm{sgn}(k_0)\frac{\mu(T)}{N} = i\frac{c_f\mathrm{sgn}(k_0)}{N}T^{2/3}H_{1/3}\left(\frac{|k_0|-\pi T\mathrm{sgn}(k_0)}{2\pi T}\right)+i\mathrm{sgn}(k_0)\frac{\mu(T)}{N},
\end{align}
which gives the expression for the fermion propagator in the main text.

\section{Wightman functions}
\label{wight}
The Wightman function for two operators $A,B$ of concern to us is
\begin{align}
&G^W_{AB}(x,t) = \mathrm{Tr}[e^{-\beta H}A(x,t)B(0,i\beta/2)] \nonumber \\
&=\sum_{nm}\langle E_n|B(0)|E_m\rangle\langle E_m|A(x,0)|E_n\rangle e^{-\beta E_n}e^{-\beta(E_m-E_n)}e^{-i(E_n-E_m)(t-i\beta/2)} \nonumber \\
&=\sum_{nm}\langle E_n|B(0)|E_m\rangle\langle E_m|A(x,0)|E_n\rangle e^{-\beta E_n}e^{-\beta(E_m-E_n)/2}e^{-i(E_n-E_m)t}. 
\end{align}
\begin{align}
&G^W_{AB}(k,\omega) = 2\pi\sum_{nm}\langle E_n|B(0)|E_m\rangle\langle E_m|\int d^dx A(x,0)e^{-ikx}|E_n\rangle e^{-\beta E_n}e^{-\beta(E_m-E_n)/2}\delta(\omega-(E_n-E_m)) \nonumber \\
&=2\pi\sum_{nm}\langle E_n|B(0)|E_m\rangle\langle E_m|\int d^dx A(x,0)e^{-ikx}|E_n\rangle e^{-\beta E_n}\delta(\omega-(E_n-E_m))\frac{e^{\beta(E_n-E_m)}\mp1}{e^{\beta(E_n-E_m)/2}\mp e^{-\beta(E_n-E_m)/2}},
\end{align}
where the $-$ sign is for bosonic operators and $+$ sign is for fermionic operators. Using the definition of the spectral function
\beq
S_{AB}(k,\omega) = 2\pi\sum_{nm}\langle E_n|B(0)|E_m\rangle\langle E_m|\int d^dx A(x,0)e^{-ikx}|E_n\rangle e^{-\beta E_n}\delta(\omega-(E_n-E_m))(e^{\beta(E_n-E_m)}\mp1),
\eeq
we have
\begin{align}
&G^W_{AB}(k,\omega) = \frac{S_{AB}(k,\omega)}{2\sinh\frac{\beta\omega}{2}}~~(\mathrm{bosons}), \nonumber \\
&G^W_{AB}(k,\omega) = \frac{S_{AB}(k,\omega)}{2\cosh\frac{\beta\omega}{2}}~~(\mathrm{fermions}).
\end{align}

\section{Higher order corrections}
\label{hoc}
We consider the corrections to the ladder series of the main text coming from diagrams with crossed rungs. We show that certain diagrams with crossed boson rungs vanish, and that diagrams with crossed fermion rungs contribute to $\lambda_L$ at higher orders in $1/N$. 
\begin{figure}
\begin{center}
\includegraphics[height=1.0in]{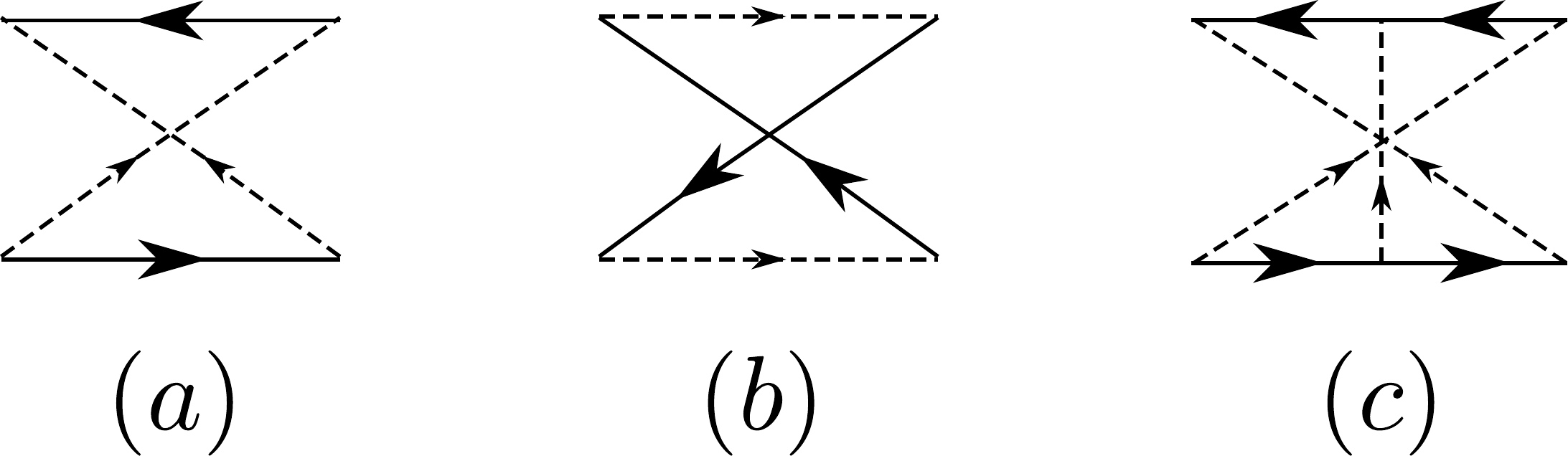}
\end{center}
\caption{(a), (b) The two simplest crossed ladder insertions in the Bethe-Salpeter equation. The first vanishes, and the second contributes to $\lambda_L$ at $\mathcal{O}(1/N)$. (c) A higher-order ``maximally crossed" diagram with boson rungs. Diagrams of this type also vanish for the same reason as (a).}
\label{crl}
\end{figure}

There are two simple types of crossed ladder insertions in the Bethe-Salpeter equation. The first is shown in Fig.~\ref{crl}a and is given by
\beq
I_1(k,k^\prime,\omega) = e^4 \int \frac{d^3k_1}{(2\pi)^3}D^W(k-k_1)D^W(k_1-k^\prime)G^R(k_1)G^{R\ast}(k+k^\prime-k_1-\omega).
\eeq
The integral over $k_{1x}$ vanishes as the $D^W$'s do not depend on $k_{1x}$ and $G^R(k_1)G^{R\ast}(k+k^\prime-k_1-\omega)$ has two simple poles both in the upper half-plane for the $k_{1x}$ integration. Thus this insertion contributes nothing. Other ``maximally crossed" diagrams of the same type (Fig.~\ref{crl}c) also vanish for exactly the same reason.

The insertion of Fig.~\ref{crl}b does not vanish. However, unlike the third diagram on the right hand side of Fig.~\ref{bs}, the flavor indices on the two sides of the insertion are not decoupled. Thus, there is no factor of $N$ enhancement from an additional sum over flavors, and this insertion is smaller by a factor of $1/N$ (The integrals for this insertion are similar to the integrals for the ``Box" insertion considered in the main text and do not contain any IR divergences that enhance its value by factors of $N$).

Finally, we must mention that, due to the uncontrolledness of the large $N$ expansion, there will be more complicated higher-loop insertions that, although naively down powers of $N$, will end up contributing at the same order as the diagrams we considered in the main text. We do not know how to systematically resum these kinds of diagrams in general, but the numerical values of these higher loop diagrams might be significantly smaller than the ones already considered~\cite{Metlitski2010}.

\section{Numerical methods}
\label{nume}
\begin{figure}
\begin{center}
\includegraphics[width=6in]{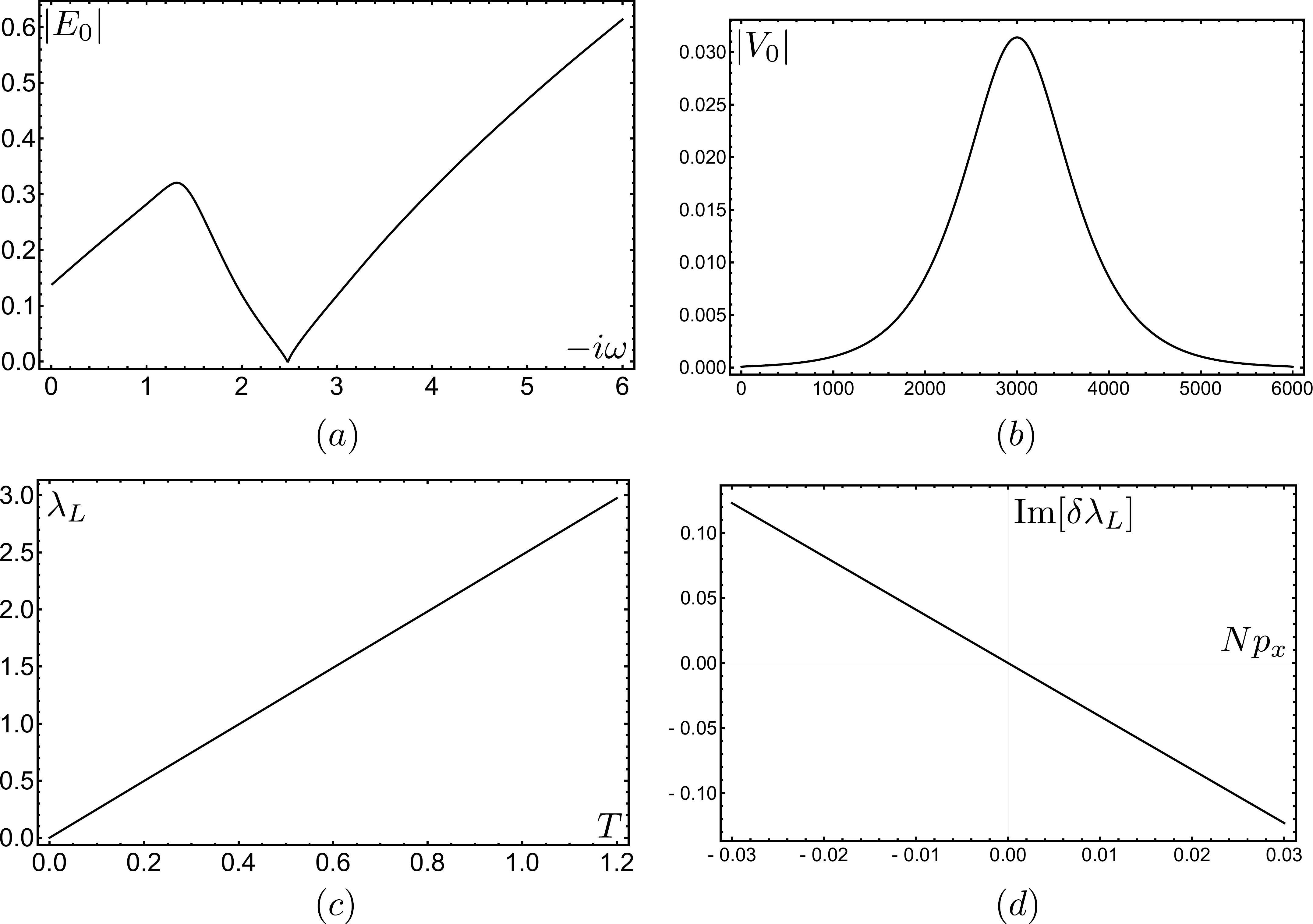}
\end{center}
\caption{(a) Plot of the magnitude of the smallest eigenvalue for $\omega$ on the positive imaginary axis for $T=1.0$. (b) Plot of the magnitude of the entries of the corresponding eigenvector when $-i\omega=\lambda_L$. (c) Plot of $\lambda_L$ vs $T$. (d) Plot of $\mathrm{Im}[\delta\lambda_L]$ vs $Np_x$. The value of $\mathrm{Re}[\delta\lambda_L]\sim (Np_x)^2$ is very small when $Np_x$ is small. This real part does not control the speed $v_{B\perp}$ at which the wave pulse of Eq.~(\ref{wave}) travels, but will lead to the broadening of the pulse as it travels (see below). For all these figures, $k_0\in[-15,15]$, $m=0.02$, the step size $dk_0=0.005$ and $e=1.0$.}
\label{nm}
\end{figure}

Numerically, it is easier to solve Eq.~(\ref{iefinal}) keeping the IR divergent term explicit.
\begin{align}
&e^2\int \frac{dk_0^\prime dk_y^\prime}{(2\pi)^2}|k_y^\prime|\frac{\tilde{f}(\omega,k_0^\prime)}{(|k_y^\prime|^3+m^2)^2+c_b^2(k_0^\prime-k_0)^2}\frac{c_b(k_0^\prime-k_0)}{\sinh\frac{\beta(k_0^\prime-k_0)}{2}} \nonumber \\
&+ \frac{e^4}{24\pi\sqrt{3}c_b^{4/3}}\int \frac{dk^\prime_0dk_{10}}{2\pi}\frac{k_{10} \left((-i k_{10})^{1/3}-(i (k_{10}-\omega ))^{1/3}\right)}{(-i k_{10})^{4/3} (i (k_{10}-\omega ))^{1/3} (2 k_{10}-\omega)}\frac{\tilde{f}(\omega,k^\prime_0)}{\cosh\frac{k_0-k_{10}}{2T}\cosh\frac{k^\prime_0-k_{10}}{2T}} \nonumber \\
&=\left[c_fT^{2/3}\left(H_{1/3}\left(-\frac{i k_0 +\pi T}{2\pi T}\right)+H_{1/3}\left(-\frac{i (\omega-k_0) +\pi T}{2\pi T}\right)\right)+2\mu(T)\right]\tilde{f}(\omega,k_0).
\label{iefinaln}
\end{align}
We keep $m$ finite but small, such that $m^2\ll T$ and $m^{2/3}\ll T$. The integration over $k_y^\prime$ is done numerically. The integration over $k_0^\prime$ then is discretized as a matrix multiplication, and the equation is brought to a form $M(\omega)\tilde{f}(\omega)=0$. For a given $\omega$ on the positive imaginary axis, we find the eigenvalue of $M$ with the smallest magnitude, which is easier to do than diagonalizing the entire matrix. We then use the Newton-Raphson method to find values of $\omega$ on the positive imaginary axis for which the smallest eigenvalue of $M$ is zero or nearly zero within a small tolerance. A plot of the magnitude of the smallest eigenvalue as a function of $-i\omega$ is shown in Fig.~\ref{nm}(a). We see that there is one zero for $-i\omega>0$, which gives the value of $\lambda_L$. The corresponding eigenvector is shown in Fig.~\ref{nm}(b). A plot of $\lambda_L$ vs $T$ is shown in Fig.~\ref{nm}(c).

For the butterfly velocity, we solve Eq.~(\ref{iefinalb}) using the same technique as in the above. Now $\lambda_L$ is no longer purely real when $Np_x\neq0$, and we numerically find $\frac{\delta\lambda_L}{\delta Np_x}$ for small $Np_x$ using the slope of Fig.~\ref{nm}(d), leading to the result in the main text. In order to determine the function $g(t,Np_x)$ that controls the shape of the wave pulse in Eq.~(\ref{wave}) of the main text, we need to numerically find $\delta\lambda_L$ to higher orders in $Np_x$. Up to second order we obtain $\delta\lambda_L/T \approx -4.10(iNp_x/(e^{4/3}T^{2/3})) - 2.74(N^2p_x^2/(e^{8/3}T^{4/3}))$. This gives
\beq
g(t,Np_x) \sim e^{-D^f_\perp p_x^2t},~~\int dy~f(t,x) \sim \frac{1}{\sqrt{tD^f_\perp}} e^{\lambda_L t}e^{-\frac{(x-v_{B\perp}t)^2}{4D^f_\perp t}},~~D^f_\perp \approx 2.74 \frac{N^2}{e^{8/3}T^{1/3}}\frac{v_F^{10/3}}{\gamma^{2/3}}
\eeq
when factors of $v_F$ and $\gamma$ are restored. The quantity $D^f_\perp$ has the dimensions and scaling of a diffusion constant such as $D^E_\perp$. However, we are unable to make any comments as to whether any special relation exists between $D^E_\perp$ and $D^f_\perp$.

\section{Specific heat and thermal conductivity}
\label{shtc}
The expression for the free energy may be rewritten as a contour integral, keeping in mind the branch cuts in the fermion propagators along the real frequency axis
\begin{align}
&F = \frac{N}{2\pi i}\int_{-\infty}^{\infty}\frac{dz}{e^{z/T}+1}\int\frac{d^2k}{(2\pi)^2}(\ln G^{-1}(z^+,k)-\ln G^{-1}(z^-,k)),~~G^{-1}(z^\pm,k)=k_x+k_y^2 \mp \frac{i\tilde{c}_f}{N}(\mp iz)^{2/3}, \nonumber \\
&=  -\frac{N}{\pi}\int\frac{dk_y}{2\pi}\int_{-\infty}^{\infty}\frac{dz}{e^{z/T}+1}\int \frac{dk_x}{2\pi}\tan^{-1}\left(\frac{\tilde{c}_f|z|^{2/3}/(2N)}{k_x-(\tilde{c}_f\sqrt{3}/(2N))\mathrm{sgn}(z)|z|^{2/3}}\right) \nonumber \\
&-N\int\frac{dk_y}{2\pi}\int_{-\infty}^{\infty}\frac{dz}{e^{z/T}+1}\int_{-\Lambda}^{(\tilde{c}_f\sqrt{3}/(2N))\mathrm{sgn}(z)|z|^{2/3}}\frac{dk_x}{2\pi},
\end{align}
where we shifted $k_x\rightarrow k_x-k_y^2$ to eliminitate $k_y$ from the integral and $\Lambda$ is some large cutoff. The $k_x$ integral over the $\tan^{-1}$ vanishes. Keeping only finite terms (which obey the quantum critical scaling),
\beq
F = -\tilde{c}_f\sqrt{3}\int \frac{dk_y}{2\pi}\int_0^\infty\frac{dz}{2\pi}\frac{z^{2/3}}{e^{z/T}+1}.
\eeq  
Evaluating this integral and differentiating with respect to $T$ gives the expression for $C_V$ in the main text.

We now turn to the computation of the energy current correlator required to determine $\bar{\kappa}_{\perp}$. The contribution which includes the resummation of the one-loop self energy corrections is
\begin{align}
&\langle J^E_\perp J^E_\perp\rangle(iq_0) = N\int\frac{d^2k}{(2\pi)^2}T\sum_{k_0}\tilde{G}(k)\tilde{G}(k+q_0)\left(k_0+\frac{q_0}{2}\right)^2 \nonumber \\
&=\frac{N^2}{2\tilde{c}_f}\int\frac{dk_y}{2\pi}T\sum_{k_0}\frac{\left(k_0+\frac{q_0}{2}\right)^2|\Theta(k_0)-\Theta(k_0+q_0)|}{|k_0|^{2/3}+|k_0+q_0|^{2/3}} \nonumber \\
&=\frac{N^2}{\tilde{c}_f}\int\frac{dk_y}{2\pi}T\sum_{k_0=\{-|q_0|\}}^{-\pi T}\frac{\left(k_0+\frac{|q_0|}{2}\right)^2}{(-k_0)^{2/3}+(k_0+|q_0|)^{2/3}}.
\end{align}
Where by $\{-|q_0|\}$ we mean the first fermionic Matsubara frequency above the bosonic Matsubara frequency $-|q_0|$. The sum can be converted into a (suitably regularized) contour integral
\begin{align}
&\langle J^E_\perp J^E_\perp\rangle(iq_0) = \frac{N^2T^{7/3}}{\tilde{c}_f}\int\frac{dk_y}{2\pi}\Bigg[\int_0^\infty\frac{dz}{2\pi i}\frac{1}{e^z+1}\left(\frac{\left(-iz+\frac{|q_0|}{2T}\right)^2}{(iz)^{2/3}+\left(\frac{|q_0|}{T}-iz\right)^{2/3}}-\frac{\left(iz+\frac{|q_0|}{2T}\right)^2}{(-iz)^{2/3}+\left(\frac{|q_0|}{T}+iz\right)^{2/3}}\right) \nonumber \\
&+\int_{-\infty}^0\frac{dz}{2\pi i}\Bigg\{\frac{1}{e^z+1}\left(\frac{\left(-iz+\frac{|q_0|}{2T}\right)^2}{(iz)^{2/3}+\left(\frac{|q_0|}{T}-iz\right)^{2/3}}-\frac{\left(iz+\frac{|q_0|}{2T}\right)^2}{(-iz)^{2/3}+\left(\frac{|q_0|}{T}+iz\right)^{2/3}}\right) \nonumber \\
&-\left(\frac{i\left(z(-iz)^{1/3}+z(iz)^{1/3}\right)}{9\left((-iz)^{2/3}+(iz)^{2/3}\right)^3}\frac{q_0^2}{T^2}-\frac{4iz}{3\left((-iz)^{2/3}+(iz)^{2/3}\right)}\frac{|q_0|}{T}\right)\Bigg\}\Bigg].
\end{align}
These integrals must be done numerically, and it is then easily seen that they reproduce the sum correctly at bosonic Matsubara $q_0$. When $q_0\rightarrow0$, we find that
\beq
\langle J^E_\perp J^E_\perp\rangle(iq_0) \approx -0.28\frac{N^2T^{4/3}}{\tilde{c}_f}|q_0|\int\frac{dk_y}{2\pi},
\eeq
which yields the result in the main text after analytic continuation.

For the conductivity $\alpha_\perp$, we have the charge current
\beq
J_\perp(iq_0) = \int\frac{d^3k}{(2\pi)^3}\frac{\partial \epsilon_k}{\partial k_x}\psi^\dagger(k+q_0)\psi(k) = \int\frac{d^3k}{(2\pi)^3}\psi^\dagger(k+q_0)\psi(k).
\eeq
Then
\begin{align}
&\langle J^E_\perp J_\perp\rangle(iq_0) = iN\int\frac{d^2k}{(2\pi)^2}T\sum_{k_0}\tilde{G}(k)\tilde{G}(k+q_0)\left(k_0+\frac{q_0}{2}\right) \nonumber \\
&=\frac{N^2}{2\tilde{c}_f}\int\frac{dk_y}{2\pi}T\sum_{k_0}\frac{\left(k_0+\frac{q_0}{2}\right)|\Theta(k_0)-\Theta(k_0+q_0)|}{|k_0|^{2/3}+|k_0+q_0|^{2/3}} = 0,~~(q_0=2n_q\pi T)
\end{align}
and hence $\alpha_\perp$ vanishes in our approximation.

The momentum integrals in the simple two-loop vertex correction to these contributions were considered in Ref.~\cite{Metlitski2010} for the higher-loop renormalizations of the boson propagator. They found that the momentum integrals in the vertex correction vanish, owing to the obtainment of terms with denominators posessing poles on the same side of the real axis.
 
\bibliography{scramble}

\end{document}